# An Enhanced "Flux-Corrected Transport"-Based Plasmasphere Refilling Model

**Jaden Fitzpatrick**[1], **Kausik Chatterjee**[1,2,3], **and Naomi Maruyama**[1]

[1]Laboratory for Atmospheric and Space Physics, University of Colorado Boulder, Boulder, CO, USA, [2]IB3 Global Solutions, Washington, DC, USA, [3]University of New Mexico, Albuquerque, NM, USA

**Key Points:**

- A multi-ion, two-stream, hydrodynamic flux-corrected transport model is enhanced to solve electron temperature self-consistently
- Incorporating temperature spatiotemporal variability in the model improves the characterization of two-stage plasmasphere refilling
- Each ion species contributes to the two-stage process in a unique manner

**Correspondence to:**

J. Fitzpatrick,
jaden.fitzpatrick@lasp.colorado.edu

**Citation:**

Fitzpatrick, J., Chatterjee, K., & Maruyama, N. (2026). An enhanced "flux-corrected transport"-based Plasmasphere refilling model. *Journal of Geophysical Research: Space Physics*, *131*, e2025JA034915. https://doi.org/10.1029/2025JA034915

Received 1 DEC 2025
Accepted 12 APR 2026

**Author Contributions:**

**Conceptualization:** Jaden Fitzpatrick, Kausik Chatterjee, Naomi Maruyama
**Data curation:** Jaden Fitzpatrick
**Formal analysis:** Jaden Fitzpatrick
**Funding acquisition:** Naomi Maruyama
**Investigation:** Jaden Fitzpatrick, Kausik Chatterjee, Naomi Maruyama
**Methodology:** Jaden Fitzpatrick, Kausik Chatterjee, Naomi Maruyama
**Project administration:** Naomi Maruyama
**Resources:** Naomi Maruyama
**Software:** Kausik Chatterjee
**Supervision:** Naomi Maruyama
**Validation:** Jaden Fitzpatrick, Kausik Chatterjee, Naomi Maruyama
**Visualization:** Jaden Fitzpatrick
**Writing – original draft:** Jaden Fitzpatrick, Kausik Chatterjee, Naomi Maruyama
**Writing – review & editing:** Jaden Fitzpatrick, Naomi Maruyama

## Abstract

A previously developed multi-ion, two-stream Flux-Corrected Transport (FCT) hydrodynamic model for plasmasphere refilling has been extended to incorporate self-consistent electron temperature evolution. The past assumption of a constant temperature along the modeled flux tube has been replaced by solving the electron energy equation, permitting spatially and temporally varying temperature. This improvement provides a more physically complete representation of the pressure and ambipolar electric-field gradients that influence ion transport. The extended model allows us to investigate two-stage refilling behavior established by prior observations and simulations. The model continues to reproduce the expected dominance of $H^+$, enhanced early-time $O^+$ contributions, and the coupling between $H^+$ and $He^+$ through the ambipolar electric field during the transition between stages. Sensitivity experiments with modified initial ion concentrations, including cases representing seasonal effects, highlight the distinct roles of each ion species in shaping the refilling trajectory. Comparisons across L-shells 3 and 4 further confirm the robustness of the model framework for future extension to three-dimensional geometries. Overall, by incorporating more realistic temperature variations, this enhanced model strengthens the physical understanding for interpreting complex multi-ion transport processes during plasmasphere recovery following geomagnetic storms.

## Plain Language Summary

The plasmasphere hosts cold, dense plasma in a torus-shape surrounding the earth. Solar storms cause most of the plasmasphere's mass to erode away. The ionosphere is the region below the plasmasphere that supplies plasma to replace what was lost from a solar storm via plasmasphere refilling. A model of plasmasphere refilling that considers two streams of multiple ions from the ionosphere now solves the electron energy equation. This allows temperature to depend on space and time instead of the previous nonphysical assumption of constant temperature. The extended model allows us to investigate two-stage refilling behavior established by prior observations and simulations. The impact of each ion on each stage of refilling was analyzed and further isolated by simulating refilling events with modified initial ion concentrations, including those reflecting seasonal variations. Comparisons across altitude further confirm the robustness of the model framework for future extension to three-dimensional geometries. Overall, by incorporating more realistic temperature variations, this enhanced model strengthens the physical understanding for interpreting the complex transport of ions during plasmasphere recovery following solar storms.

## 1. Introduction

The plasmasphere (Banks et al., 1971; Carpenter & Park, 1973; Darrouzet et al., 2009; Goldstein et al., 2002; Gringauz, 1962; Millan & Thorne, 2007; Pezzopane et al., 2019; Sandel et al., 2003) is the innermost and most massive region of the magnetosphere and resides above the ionosphere. A geomagnetic storm causes the sunward movement of plasma from the plasmasphere, eliminating its outer layers, and thus creates a pressure gradient between the ionosphere and depleted plasmasphere. Consequently, ionospheric plasma is forced to flow upward along the flux tubes that bridge the ionosphere and plasmasphere, initiating the refilling process.

Attempts have been made to observe plasmasphere refilling, starting with ESA's GEOS-2 satellite (Sojka & Wrenn, 1985; Song et al., 1988) and the Los Alamos National Laboratory (LANL) Magnetospheric Plasma Analyzers (MPAs) (Lawrence et al., 1999; Su et al., 2001) collecting data at geosynchronous orbit. Refilling rates and trends were obtained, but the intricate structure of the plasmasphere and its lack of corotation (Burch et al., 2004) complicated the interpretation of the data. At the time of the previously stated studies, there were no indications that refilling was a multi-stage process. It was not until analyses such as Lawrence et al. (1999); Su et al. (2001) that refilling was suggested to be composed of two stages. Each stage exhibited a different rate of refilling, with the second stage having a significantly greater refilling rate than the first. A more recent study of Van Allen Probe data (Bishop et al., 2025) also observed two-stage plasmasphere refilling, but only in approximately 40% of the examined refilling events. The study concluded that the local plasma environment is the most probable determinant for one-versus two-stage refilling.

Hydrodynamic and semikinetic models of plasmasphere refilling have also been developed to compare with observation and provide more physical insight as to what processes influence refilling. The first models (Banks et al., 1971; Khazanov et al., 1984; Singh et al., 1986) were hydrodynamic and treated the plasma as a single fluid traveling along a flux tube, resulting in equatorial shocks that were later demonstrated to be numerical artifacts (Rasmussen & Schunk, 1988). The two-stream approach (Guiter et al., 1995; Rasmussen & Schunk, 1988) that followed eliminated any nonphysical shocks because distinguishing the north- and south-bound streams accurately models the interpenetration of plasma originating from either hemisphere. Wilson et al. (1992); Lin et al. (1992) attempted a semikinetic framework, which is optimal for the low concentrations early in refilling, and again demonstrated a two-stage refilling process. More recent models (Chatterjee & Schunk, 2020b; Krall & Huba, 2013; Liemohn et al., 1999) enhancing pre-existing works or proposing new means of analysis continue to expose plasmasphere refilling dynamics.

Previous multi-ion models using the Flux-Corrected Transport (FCT) method (Boris & Book, 1976) - particularly well-suited to handle problems with shocks and discontinuities - reproduced many features of refilling. Still, they assumed a spatially uniform and constant temperature along the flux tube (Chatterjee & Schunk, 2019, 2020a). Under this assumption, the pressure gradient and ambipolar electric field are determined entirely by density gradients.

The objective of this work is to extend the existing multi-ion, two-stream FCT hydrodynamic model, first introduced in Chatterjee and Schunk (2019); Chatterjee and Schunk (2020a), by incorporating the electron energy equation. Therefore, plasma temperatures are allowed to vary self-consistently in space and time during refilling evolution. This extension enables investigation of how temperature evolution modifies the pressure gradient and ambipolar electric field, influences the acceleration of different ion species, and alters the timing and structure of early and late refilling stages. As shown in the following sections, temperature-coupled dynamics introduce new physical pathways that help explain observed variability in two-stage refilling behavior and provide a more complete description of multi-ion transport in the depleted plasmasphere.

## 2. Model Development

The governing plasma transport equations solved in this study follow the multi-ion, two-stream hydrodynamic formulation presented in Chatterjee and Schunk (2019); Chatterjee and Schunk (2020a). Those works provide the full expressions for the continuity and momentum equations, the numerical FCT implementation, and the treatment of ion–ion and ion–neutral collisions. Accordingly, only the modifications introduced by the temperature-coupled extension are summarized here.

### 2.1. Electron Temperature Evolution

The key advancement in the present model is the inclusion of a time- and space-dependent electron temperature, obtained by solving the one-dimensional heat conduction equation of Khazanov et al. (1992):

$$\frac{3}{2}k\,n_e\,A(s)\,\frac{\partial T_e}{\partial t} = \frac{\partial}{\partial s}\left[A(s)\,K_e\,T_e^{5/2}\frac{\partial T_e}{\partial s}\right] + A(s)\,Q_e, \tag{1}$$

where $n_e$ is the electron density obtained from quasi-neutrality, $A(s)$ is the magnetic-flux-tube cross-sectional area, $K_e$ is the electron thermal conductivity, and $Q_e$ is the electron heating rate. This is the plasmasphere solution to Khazanov et al. (1992)'s general heat conductivity equation where the altitudes are assumed to be high enough to ignore loss terms. An expression for $K_e$ is given in Schunk and Walker (1970) derived by Banks (1966a):

$$K_e = \frac{7.7 \times 10^5}{1 + 3.22 \times 10^4 \frac{T_e^2}{n_e} \sum_j N_j Q_j} \text{eVcm}^{-1}\text{sec}^{-1}\text{K}^{-1}. \quad (2)$$

In Equation 2 above, the summation in the denominator is over all neutral species and the quantity $Q_j$ represents the average momentum transfer cross section, given in Banks (1966b). It should be noted that the temperature dependence of the electron thermal conductivity is expressed explicitly in Equation 1 through the $T_e^{5/2}$ factor. Accordingly, the coefficient $K_e$ in Equation 2 represents the remaining collisional contribution following the formulation of Schunk and Walker (1970), so that the full heat conduction term remains consistent with the expression used by Khazanov et al. (1992).

At plasmaspheric altitudes and particularly during the early stages of refilling, the plasma can become weakly collisional as the electron density decreases and the electron mean free path increases. While the present study adopts the classical heat-conduction formulation for consistency with previous hydrodynamic plasmasphere models, this approximation may overestimate the conductive heat flux in regions where the electron mean free path becomes comparable to the characteristic scale length of the temperature gradient. Future model developments will investigate modified or reduced heat-conductivity formulations (e.g., Denton and Bailey (2001)) to better represent electron heat transport in weakly collisional plasmaspheric conditions.

Following Khazanov et al. (1992), radiative and inelastic cooling are neglected at plasmaspheric altitudes. Equation 1 is advanced using the Crank–Nicolson method (Smith, 1985), and its solution is updated self-consistently at each plasma time step. The ion temperature for all ion species is assumed equal to the local electron temperature.

### 2.2. Modified Ambipolar Electric Field

In the earlier constant-temperature model (Chatterjee & Schunk, 2019), the electron pressure gradient reduced to a function of density alone. With spatially varying electron temperature, the ambipolar electric field becomes

$$E_{\|}(s) = -\frac{1}{e\, n_e}\frac{\partial}{\partial s}(n_e k T_e), \quad (3)$$

which naturally incorporates both density and temperature gradients. This expression is substituted directly into the ion momentum equations for each stream, replacing the constant-temperature form used in previous implementations. Because $T_e(s,t)$ evolves dynamically, the ambipolar electric field responds immediately to changes in the local thermal structure, thereby modifying the early-time acceleration of $H^+$, $He^+$, and $O^+$.

### 2.3. Coupled System and Numerical Integration

At each time step, the updated electron temperature from Equation 1 is used to compute the electron pressure gradient and ambipolar field via Equation 3. These quantities are then provided to the FCT solver, which advances the continuity and momentum equations for each ion species. This iterative coupling continues throughout the refilling period. The new formulation reduces the need for prescribed temperature profiles and allows the plasma temperature to adjust self-consistently with density evolution, thereby producing a more realistic representation of early-time dynamics and stage transitions than earlier constant-temperature models.

The complete coupled system of equations therefore consists of: (a) multi-ion continuity and momentum equations from Chatterjee and Schunk (2019), (b) the electron-energy equation Equations 1 and 3 the generalized ambipolar electric field Equation 3. This extension enables the first investigation within the FCT hydrodynamic framework of how evolving temperature gradients and thermal conduction influence ambipolar coupling, multi-ion flow, and the timing of early and late stages of plasmasphere refilling.

### 2.4. Simulation Configuration

All simulations follow the geometric configuration described in Chatterjee and Schunk (2019), but now include variable electron temperature in space and time. The modeled field line corresponds to $L = 4$, extending from

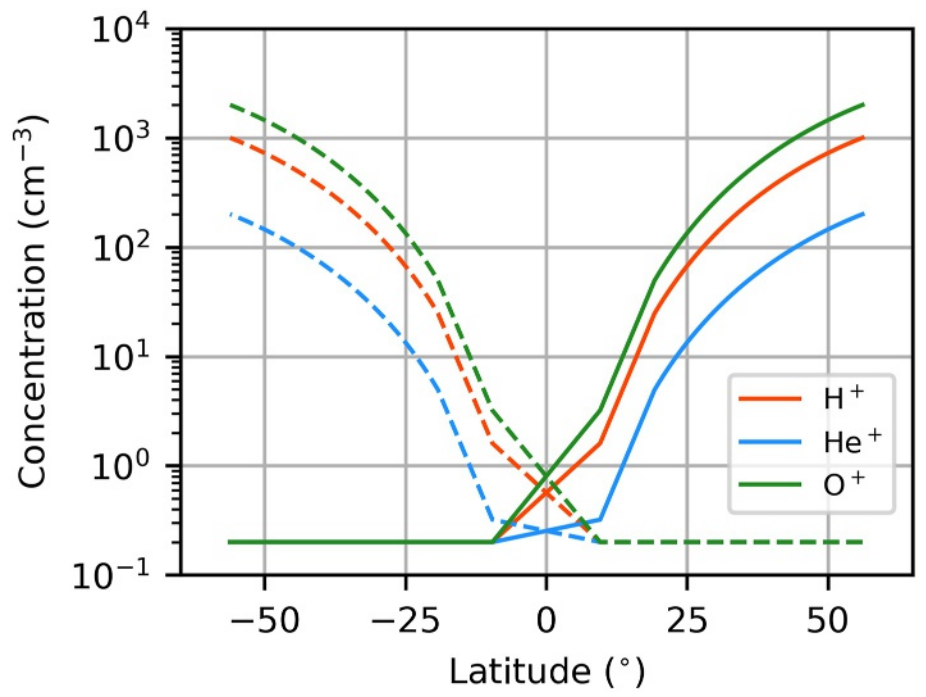

**Figure 1.** Initial concentrations for the southern hemisphere (dashed) and northern hemisphere (solid) streams assumed for a depleted plasmasphere. The maximum concentrations for each ion and hemisphere's stream are set manually, which are symmetric about the equator for this standard case. The latitudes outside a stream's designated hemisphere are set to 0.2 cm$^{-3}$ and intermediate latitudes are solved analytically.

±56° magnetic latitude to 1,500 km altitude (boundary altitude). In order to study L-dependence as shown in Figure 6, the model configuration has been modified to $L = 3$ for related simulations. Boundary temperatures at both ionospheric ends are fixed to the chosen initial value of 3560 K. The electron heating rate $Q_e$ is held at a constant value of $10^{10}$ eV cm$^{-2}$ s$^{-1}$ in space and time for each simulation. Although the constant boundary temperature and heating rates represent an idealized configuration, they allow the effects of temperature variation to be isolated and examined clearly in the present study. Variable boundary temperature and heating rates will be introduced in future work so that the effects of each improvement are distinguishable and thus indicative of their contribution and importance thereof in the refilling process.

The standard initial ion density conditions are specified as shown in Figure 1. The simulated $He^+$ abundance is sensitive to the assumed initial ion composition at the ionospheric boundaries. Because the present study focuses primarily on the dynamical effects introduced by temperature coupling, simplified initial composition profiles were adopted. Constraining the ion composition using more realistic ionospheric conditions, including coupling with ionospheric chemistry models, will be explored in future work. A suite of simulations is performed across a range of initial ion densities for each ion species that assume some proportion of the standard conditions, enabling a systematic examination of how the resulting spatiotemporal variation in temperature influences the timing and structure of multi-ion plasmasphere refilling.

## 3. Results

This section presents the effects of incorporating the electron energy equation into the multi-ion, two-stream refilling model under consideration. Because the underlying $H^+$, $He^+$, and $O^+$ transport behavior in the absence of temperature variation has been documented extensively in earlier work (Chatterjee & Schunk, 2019, 2020a), the discussion here focuses on the new physical consequences arising from spatiotemporal temperature evolution and on the sensitivity of two-stage refilling to ion composition and L-shell. The results displayed assume an initial temperature of 3560K, $L = 4$, and the initial concentrations from Figure 1 unless otherwise specified.

### 3.1. Spatiotemporal Evolution of Temperature

The newly implemented time-varying electron temperature is depicted in Figure 2 with each colored line signifying the electron temperature across latitude at its corresponding 10-min interval after refilling began. The only remanence of the initially constant temperature of 3560 K is the boundary temperature. After 10 min there is already a temperature gradient between the equator and the highest latitudes of approximately 1500 K. The temperature profile experiences minimal changes after 60 min and for the rest of refilling, as demonstrated by the equatorial temperature over time in Figures 4 and 5. Thus, it is assumed that at approximately 60 min, the temperature has reached an equilibrium. Reaching steady-state this rapidly is to be expected considering the lack of dynamicism in the terms determining the electron temperature, such as by ignoring loss-terms and assuming time- and space-independent heating rates.

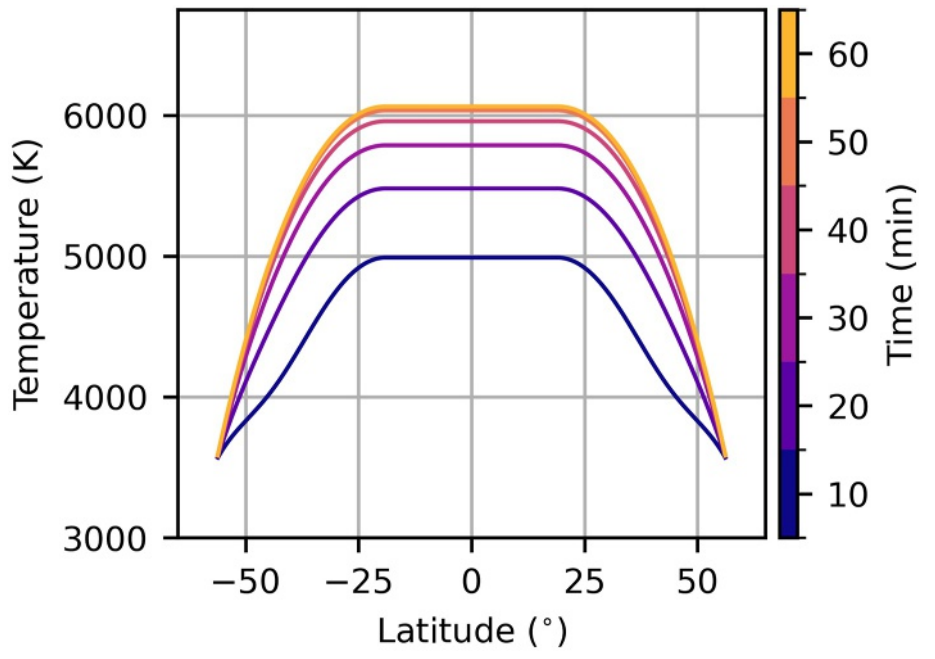

**Figure 2.** Electron temperature across latitude at 10-min intervals after a constant initial temperature of 3560 K.

### 3.2. Influence of Temperature Variability on Refilling

In the middle and bottom panels of Figure 3, the refilling profile produced by the newly enhanced model with temperature variability (T = T(s, t)) is compared against that from the previous model version where temperature is held at a constant 3560 K (T = $c$). The middle panel displays the total ion concentration (sum of $H^+$, $He^+$, and $O^+$) and shows that both model versions have similar behaviors for nearly 4 h before the constant temperature version breaks away to a higher concentration for the remainder of the simulation. From the bottom panel, it is observed that the fractional $He^+$ is much lower in the constant temperature regime and has less variability in the first 15 h of refilling than when temperature variations are taken into account. The fractional $O^+$ from each model version is extremely similar and therefore challenging to distinguish between the two in the bottom panel.

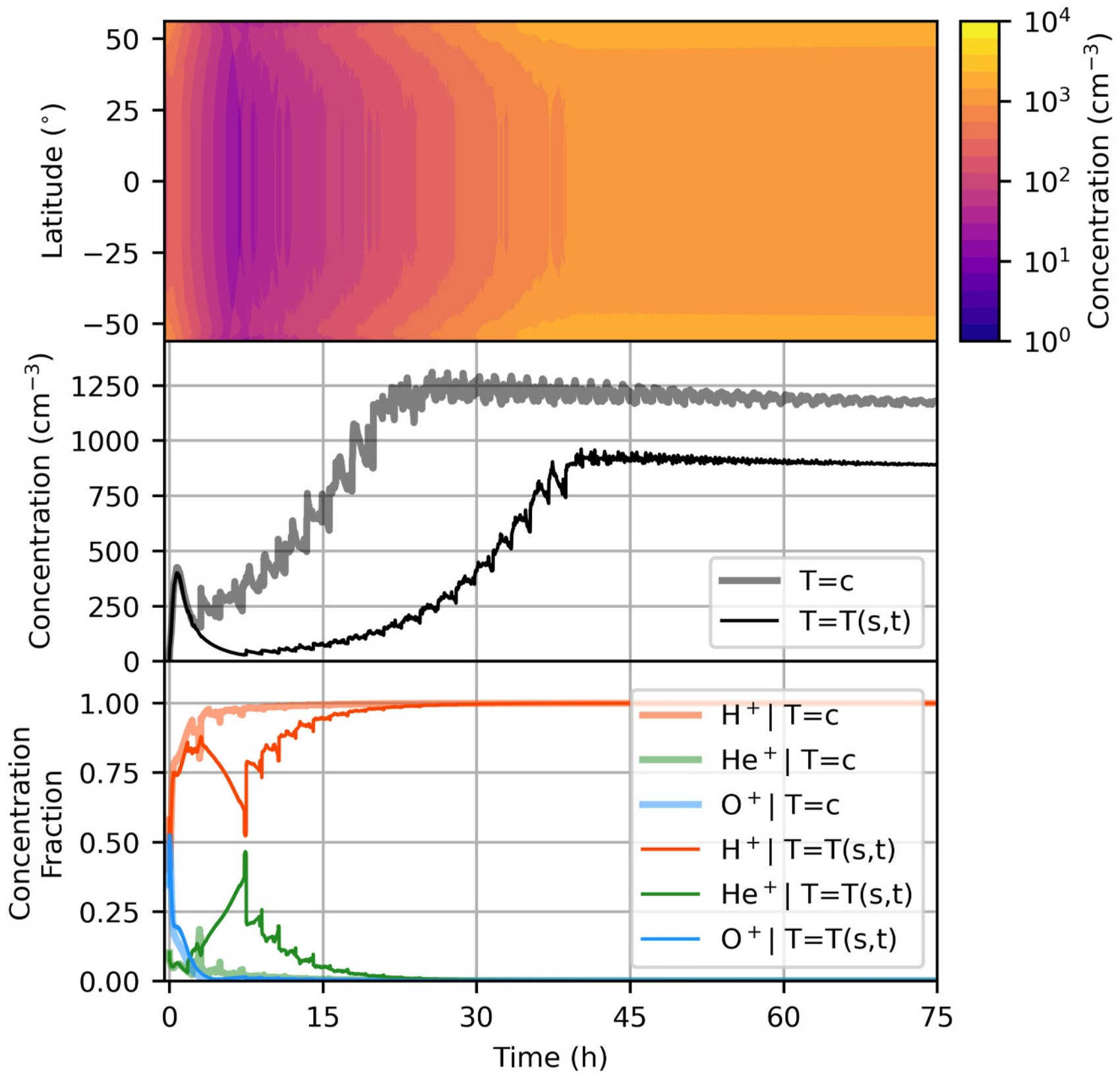

**Figure 3.** (Top) Concentration of $H^+$ as a function of latitude and time to visualize how the flux tube refills across all latitudes. (Middle) Total equatorial concentration of all ions summed ($H^+ + He^+ + O^+$) over the course of refilling before (transparent black line, T = c) and after (solid black line, T = T(s, t)) spatiotemporally varying temperature was implemented. (Bottom) The fractions of $H^+$, $He^+$, and $O^+$ out of the total ion concentration at the equator as functions of time. Similar to the middle panel, the transparent lines are from when the model assumed constant temperature (T = c), while the solid lines were produced by the model when incorporating temperature variability (T = T(s, t)).

Assuming refilling concludes once the total concentration reaches a near-constant value, it is evident that the latest model version significantly lengthens the time taken to refill and reduces the refilling rate, both of which move the modeled refilling behavior closer to observationally inferred features (Bishop et al., 2025; Gallagher et al., 2021; Lawrence et al., 1999; Sojka & Wrenn, 1985; Song et al., 1988; Su et al., 2001). Between the constant and variable temperature versions of the model, the length of refilling went from approximately 26–37h and the average refilling rate over that time from 1,224.5 cm$^{-3}$ to 573.0 cm$^{-3}$. Still, the refilling lengths derived from the newly enhanced model are minimal compared to the 5+ day-long scales reported by Bishop et al. (2025). Similarly, the gathering of refilling rates across various studies (Gallagher et al., 2021) reported rates half to two orders of magnitude smaller than those predicted by temperature variability in our model. However, enabling temperature to vary across space and time has helped the model-predicted refilling lengths and rates to approach the observed values closer than was done previously. This fact demonstrates the value in exploring the effects of temperature variability as shown by this model in its current state, and foreshadows what it will become as assumptions of constant heating rate boundary temperature are eradicated in future iterations.

Comparing each ion's equatorial concentration over time in Figure 3 reveals evidence of two-stage refilling (Bishop et al., 2025; Lawrence et al., 1999; Lin et al., 1992; Wilson et al., 1992), which was not detectable when ignoring temperature fluctuations across space and time (Chatterjee & Schunk, 2019). Early-time refilling is assumed to begin at the start of the simulation (0 h) and ends at approximately 7 h, when the $H^+$ concentration's rate of increase is significantly enhanced. Late-time refilling proceeds until approximately 37 h when a saturated concentration is reached and refilling completes. Comparing the total equatorial concentrations (summing $H^+$, $He^+$, and $O^+$) before and after each stage implies an early-time refilling rate of 90.0 $cm^{-3}$ $day^{-1}$, followed by a late-time refilling rate of 680 $cm^{-3}$ $day^{-1}$.

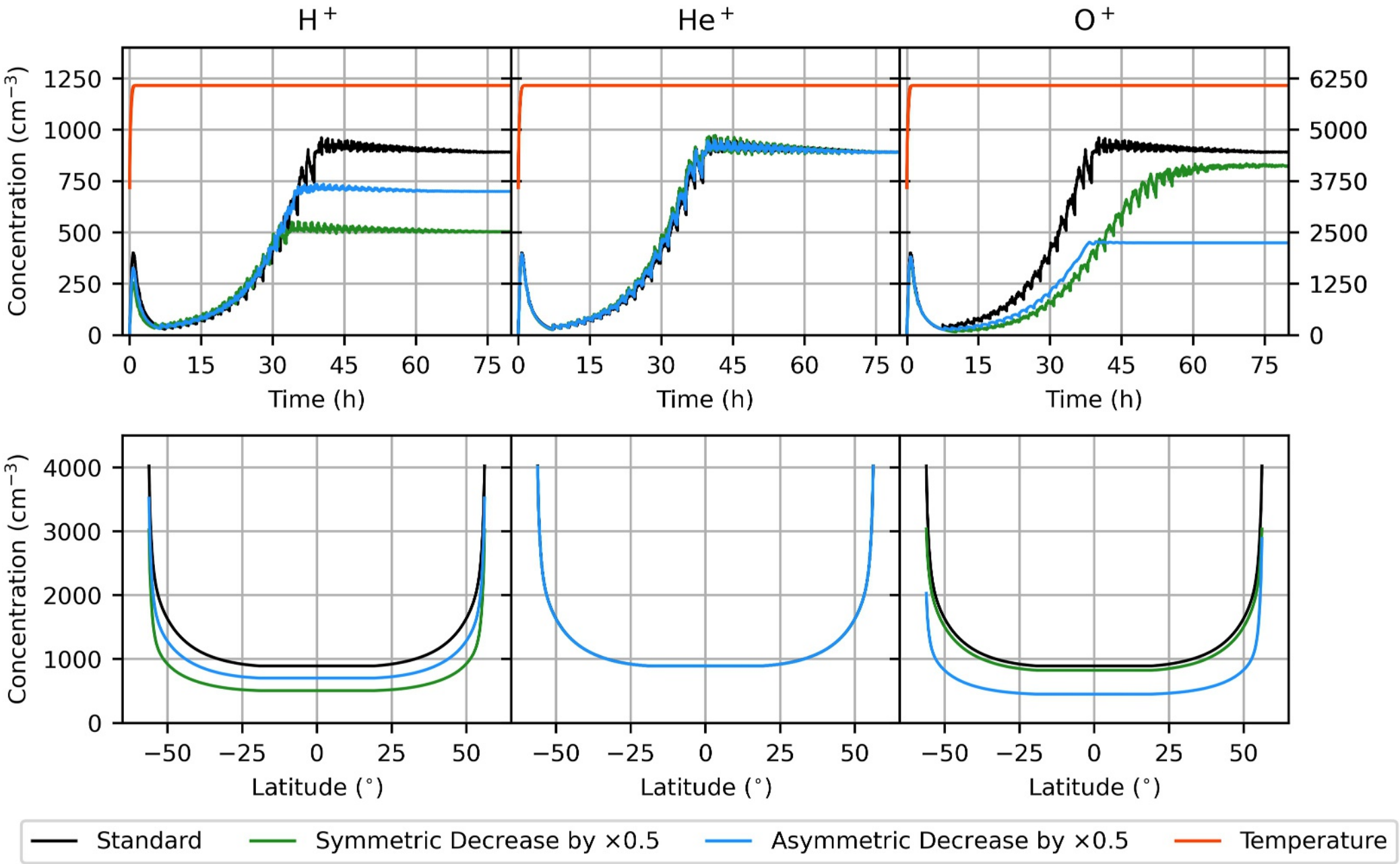

**Figure 4.** (Top) Total equatorial ion concentration (sum of $H^+$, $He^+$, and $O^+$) and equatorial electron temperature over time. The alteration of equatorial electron temperature between simulations is minimal and only at the start of refilling, which can be viewed in greater detail in Figure 5. (Bottom) Final total ion concentration across latitude. Reducing the initial value of $He^+$ in one and both hemispheres produced negligible differences, so the results of which heavily overlap.

Near the start of the early stage of refilling at approximately 1 h, a prominent peak in each of the ion species' equatorial concentrations can be observed in the middle panel of Figure 3. This appears to be contrary to other studies (Bishop et al., 2025; Lawrence et al., 1999; Lin et al., 1992; Wilson et al., 1992) where the early-time refilling rate is relatively low compared to the late-time refilling rate. However, this peak in concentration is only because of the initial transport of plasma across each stream and should not be factored into the early-time refilling rate. Figure 1 shows how each ion is described by two streams that each initially have significantly greater concentration in opposite hemispheres. The density asymmetry across latitude in each stream induces an immediate, intense outflow of ions at the start of refilling. The peak in equatorial concentration at approximately 1 h in the middle panel of Figure 3 depicts when this initial outflow crosses the equator. The following decrease in equatorial concentration until the end of early-time refilling near 7 h is from the ions in the initial flows settling at higher latitudes. It is evident from the top panel of Figure 3 that for late-time refilling to achieve higher concentration at all latitudes, the concentration at higher latitudes must increase before the concentration at low latitudes. Altogether, the first 7 h of refilling in Figure 3 should not be interpreted as a high refilling rate followed by a negative refilling rate. Rather, this time segment is better described as having large latitudinal concentration variations amidst a low refilling rate, as inferred by the slight net increase in concentration.

At the transition between the refilling stages, there is a minimum in the concentration of $H^+$ and a maximum in the concentration fraction of $He^+$. The simultaneity of these events suggests the ambipolar electric field (Equation 3) is enhancing the equatorial concentration of $He^+$ in response to the loss of $H^+$ at the equator to maintain the overall ion concentrations between the stages in local space. Since the ambipolar electric field is proportional to the temperature gradient, only the dynamically solved temperature of the current model could portray such a phenomenon.

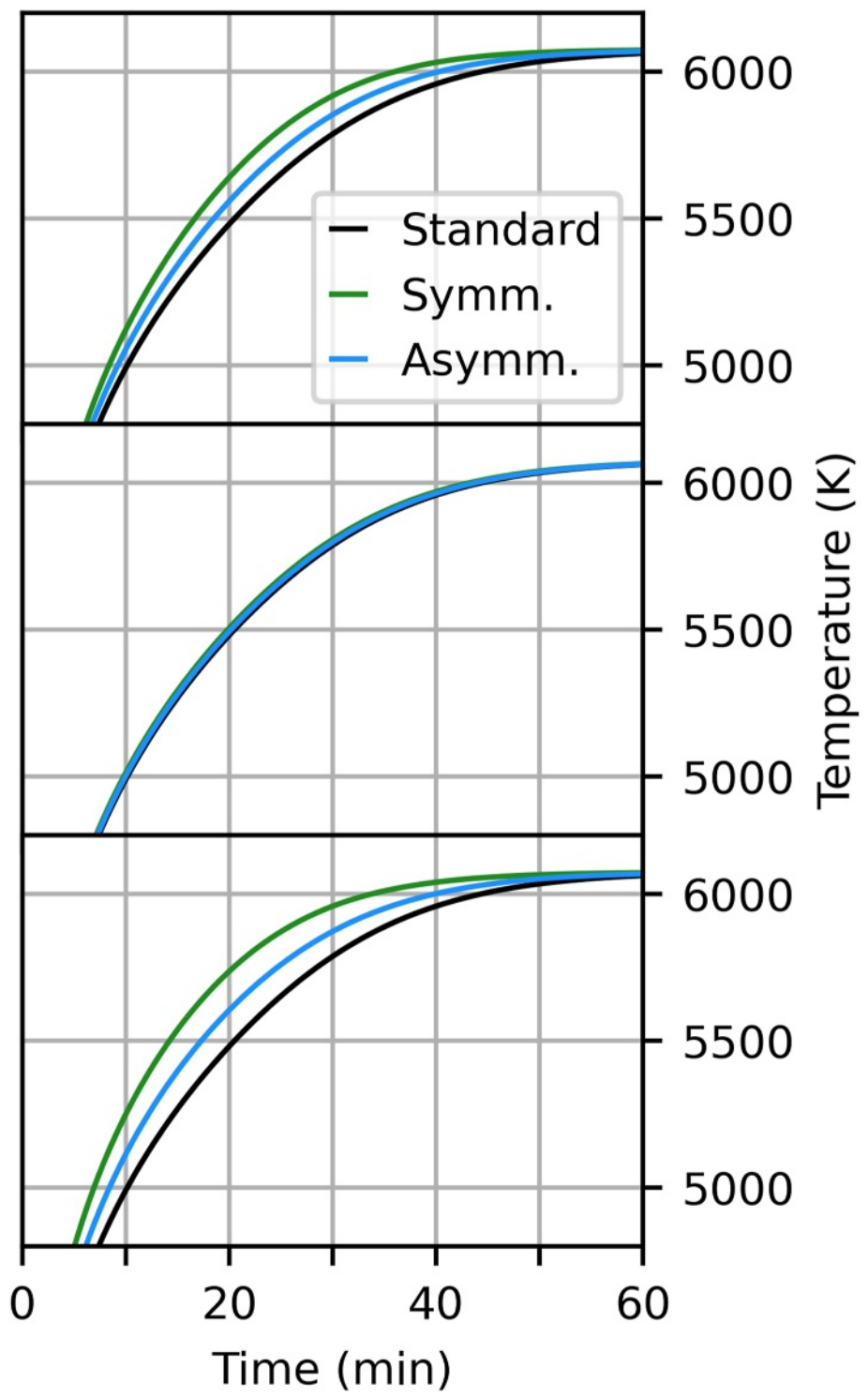

**Figure 5.** Equatorial electron temperature for the $H^+$ (top), $He^+$ (middle), and $O^+$ (bottom) simulations depicted in Figure 4. The simulations for each ion correspond to those in Figure 4 where the initial value of the ion was the Standard, decreased by a factor of 0.5 in both the northern and southern hemispheres (Symmetric), or decreased by a factor of 0.5 in only the southern hemisphere (Asymmetric).

### 3.3. Sensitivity to Initial Multi-Ion Composition

To evaluate the impacts of each $H^+$, $He^+$, and $O^+$ on two-stage refilling, Figures 4 and 5 display the results of individually reducing the initial concentration of each ion. The symmetric decrease scenario assumes the boundary concentration of the given ion in both hemispheres is reduced by half and the concentrations at other latitudes are solved to reflect the same structure as the standard initial conditions in Figure 1. The asymmetric case only decreases the initial value of an ion in the southern hemisphere alone as if the region were at or near a winter solstice.

The concentration profiles in the $H^+$ column of Figure 4 demonstrate that altering the initial $H^+$ value does not have a measurable effect on early-time refilling, but determines when and at what concentration late-time refilling concludes. The same plot in the $He^+$ column suggests that the amount of $He^+$ at the start of refilling has a negligible impact on the rest of refilling. Finally, varying the initial $O^+$ also did not influence early-time refilling. However, unlike when the $H^+$ and $He^+$ concentrations were reduced, the impacts on late-time refilling were distinguishable between the symmetric and asymmetric decrease in $O^+$. Starting with higher overall $O^+$ concentration in the asymmetric scenario led to a lower equilibrium concentration than in the symmetric case, which also reached its equilibrium concentration more gradually than any other scenario.

As expected by maintaining the initial temperature and heating rate for each situation, the final electron temperatures were all nearly identical to the 60 min profile in Figure 2 and thus were not shown. In the same vein, minimal variations occurred in the equatorial temperature versus time. The minute differences that did occur as seen in Figure 5 are a result of the temperature's dependence on the density gradient. Most of the final concentrations depict the same symmetric shape translated vertically according to the final equilibrium concentration reached at the equator seen in the top row of refilling profiles. The only exception is for the asymmetric decrease in $O^+$, where the southern hemisphere preserved its reduced concentration for the entirety of refilling.

### 3.4. L-Shell Dependence

To evaluate the model's consistency across L-value, results for $L = 3$ and $L = 4$ are compared in Figure 6, showing the summed concentration of each ion and the temperature over time. Each stage of refilling was compressed for $L = 3$ such that early-time refilling went from a duration of 7.32 to 2.31 h and late-time refilling from 32.9 to 20.0 h for $L = 4$ and $L = 3$, respectively. Concentrations reached at the end of each stage were also significantly larger for $L = 3$. The total concentration went from 27.4 to 258 $cm^{-3}$ after early-time refilling and 961 to 1,860 $cm^{-3}$ after late-time refilling when the equilibrium concentration was reached.

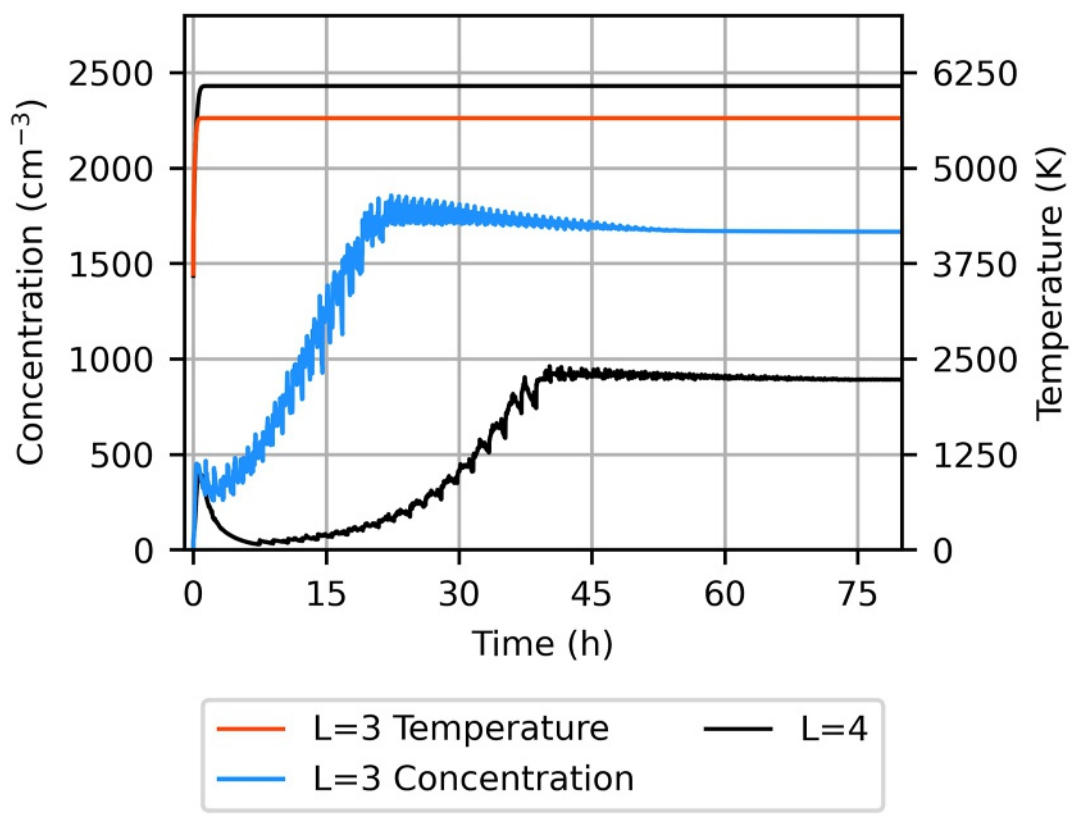

**Figure 6.** Total equatorial ion concentration (sum of $H^+$, $He^+$, and $O^+$) and equatorial temperature over time for $L = 3$ and $L = 4$.

## 4. Discussion

The temperature-coupled simulations presented here provide new physical insight into plasmasphere refilling by explicitly resolving the development of field-aligned temperature gradients and their dynamical influence on multi-ion transport, leading to two-stage refilling processes. Several key implications arise from the incorporation of the electron energy equation into the FCT, multi-ion, two-stream, hydrodynamic framework.

### 4.1. Validation of the Temperature-Coupled Model Simulations

The temperature-coupled simulations presented here reproduce the foundational characteristics of two-stream plasmasphere refilling established by prior hydrodynamic models. Although we do not include separate two $H^+$ stream plots here, the early-time refilling behavior in Figure 3 (0–7 h) is fully consistent with the "counter-streaming" behavior documented in Chatterjee (2018); Chatterjee and Schunk (2019, 2020a). In those earlier works, explicit stream-resolved velocity and density profiles (Figures 3 and 4 in Chatterjee and Schunk (2019)) demonstrated that the initial $H^+$ flows from each hemisphere are supersonic and distinct from each other. The timing of the first $H^+$ equatorial density enhancement at ~0.7 h in Figure 3 reflects the two-stream meeting at the equator. As established by Rasmussen and Schunk (1988) and later reproduced by Chatterjee (2018), the two streams interpenetrate smoothly without forming shocks, which is preserved by the present temperature-coupled version.

The early-to-late transition (after 7 h as shown in Figure 3) reflects the expected increase in Coulomb collisions as densities rise, together with the weakening of the initially strong pressure-gradient- and ambipolar-field-driven flows. The corresponding $H^+$ velocities decrease as analyzed in detail by Chatterjee (2018); Chatterjee and Schunk (2019). Increased Coulomb collisions establish a diffusive equilibrium by thermalizing the two streams, which is consistent with long-standing observational inferences that Coulomb collisions dominate the late-time refilling regime (Lawrence et al., 1999; Su et al., 2001).

### 4.2. Role of Temperature Coupling in Early-Time Refilling and the Two-Stage Process

The rapid development of a strong field-aligned temperature gradient within the first hour of refilling (Figure 2) fundamentally modifies the early-time acceleration of all ion species. Because the ambipolar electric field depends on the spatial variation of electron pressure, $\nabla(n_e k T_e)$ in Equation 3, the emergence of enhanced mid-latitude temperatures produces a substantially stronger pressure gradient than would arise under the constant-temperature assumption used in Chatterjee and Schunk (2019). The resulting increase in the ambipolar electric field leads to more vigorous initial ion acceleration and earlier peaks in the relative fractions of $He^+$ and $O^+$ than in the constant-temperature case. These enhanced early-time heavy-ion signatures are a direct consequence of the self-consistent electron temperature evolution introduced in the present model. Although the final, steady-state temperature profiles ultimately converge to similar latitudinal structures across all simulations (Figure 2), the temporal evolution of the equatorial electron temperature remains tightly coupled to the evolving ion density profiles (Figure 5). This feedback between temperature evolution and multi-ion transport influences the timing and sharpness of the transition between early- and late-time refilling, reinforcing the importance of including temperature coupling when evaluating the physical mechanisms underlying the two-stage refilling process.

### 4.3. Comparison of Two-Stage Refilling With Previous Literature

The distinct early- and late-time refilling stages depicted in the results of this model reaffirm the two-stage refilling processes in observations and other models. Observations from the Los Alamos National laboratory (LANL) Magnetospheric Plasma Analyzers (MPAs) at geosynchronous orbit by Lawrence et al. (1999); Su et al. (2001) demonstrate two refilling stages. Their data showed that early-time refilling resulted in a marginal increase in ion concentration, as does this paper's model. Lawrence et al. (1999); Su et al. (2001) suggest that while early-time refilling is dominated by wave-particle interactions, they are overtaken by Coulomb collisions for late-time refilling. A significantly enhanced refilling rate sets in until refilling completes, as is also depicted by our model. The recent results of Bishop and Blum (2024); Bishop et al. (2025) reinforce the relative refilling rates between the stages, but with data gathered from the Van Allen Probes. Figure 6 also appears to suggest a trend between length and rate of refilling of each stage across L-value. Along with Figure 5, the semikinetic model of Wilson et al. (1992); Lin et al. (1992) demonstrates that as L-value decreases, each stage of refilling shortens and experiences a greater rate of refilling. The shorter length and lower overall altitude of a flux tube with a smaller L-value would suggest these characteristics.

### 4.4. Discussion of Multi-Ion Refilling

A novel aspect of this model is that each ion's concentration is solved separately, thus allowing for the effects incurred by individual species to be analyzed. In Figure 3 is a peak in the fraction of $O^+$ near the start of early-time refilling, which has been experimentally observed (Chappell, 1982; Nosé et al., 2018). In particular, Burke et al. (2016) described the Van Allen Probe observations of the dynamics of $O^+$ ions during the 1 June 2013 storm. The reason behind this observation can be explained by the pressure gradient and the ambipolar electric field terms in the momentum conservation equation (Equations 2 and 3), which are high at the onset of refilling and more than offset the gravitational force on the $O^+$ ions. As refilling exits the first stage, the pressure gradient and electric field terms decrease, and $O^+$ is unable to reach the equatorial plane.

As pointed out previously, the transition between the two stages of refilling correlates with a peak relative abundance of $He^+$ (Figure 3). Observations have been made of an influential population of $He^+$ residing in the topside ionosphere (Chappell, 1982). Therefore, it would be reasonable for the largest presence of $He^+$ to occur during early and the start of late-time refilling in our model. Before $H^+$ concentrations dominate later on, $He^+$ is contributing significantly to the overall ion and electron concentrations. Accounting for electron temperature, per the latest model developments, permits calculating the ambipolar electric field (Equation 1). The electric field's consideration of the electron or total ion number concentrations demonstrates its role in maintaining quasi-neutrality. When $H^+$ decreases just before late-time refilling, the increase in the fraction of $He^+$ is a result of the electric field force compensating for the charge deficit incurred by losing $H^+$.

Late-time refilling was also altered by the variations in initial concentration of each ion, including when only one hemisphere's initial concentration was changed when seasonal variations were simulated. For $H^+$, reducing its overall initial concentration caused a reduction in the saturated concentration and duration to reach that concentration, regardless of how its initial concentration is distributed between the two hemispheres. Reducing $He^+$ had minimal impact, while similar to $H^+$ in that it did not matter whether $He^+$ was reduced in one or both hemispheres. However, the results of reducing $O^+$ in both hemispheres or one differed substantially with respect to the saturated density and late-time refilling length. The case for which both hemisphere's initial $O^+$ was decreased, for instance, was the only simulation that had a smoother slope when reaching the saturated concentration at approximately 60 h. This behavior more closely aligns with that seen by previous models, including Wilson et al. (1992); Singh et al. (1994). Such variability caused by slight adjustments in the initial $O^+$ may also support the notion that heavy ions play a key role in significant portions of the refilling process.

### 4.5. Model Limitations and Future Work

The conclusions drawn from the results of testing this model should consider the limitations of the model at its current state. A semikinetic model may provide a more detailed representation of the early phase of refilling at very low densities and higher altitudes because Coulomb collisions cannot thermalize at the low number densities present at these locations and times. The velocity distribution under these conditions is also not Maxwellian, but the hydrodynamic approach assumes Maxwellian velocity distributions at all times. Since early-time refilling accounts for a smaller fraction of refilling and the hydrodynamic model still shares overarching refilling characteristics with the semikinetic model from Wilson et al. (1992), the hydrodynamic approach remains suitable for describing late-time refilling.

The boundary conditions for temperature and concentration are also assumed to be stagnant, ignoring diurnal variations in the topside ionosphere and other variances associated with the recovery from a geomagnetic storm. The trends from individual tests can be applied to different segments of a given refilling event while this time dependence is not accounted for.

Temperature is still assumed to be initially constant across latitude, and the heating rate constant across latitude and time. Both confine complicated heating mechanisms to the values permitted by the model, which in reality vary across multiple length and time scales. However, the complexity of observing and accounting for individual heating mechanisms at the moment validates the need to assume a single heating rate for the sake of this analysis.

The ultimate aim for this model is to eliminate the previously mentioned simplifying assumptions. Accounting for time-dependent boundary temperature and spatiotemporally varying heating mechanisms are predicted to make the largest impact by inducing more expected refilling rates and lengths. Even without considering conditions caused by geomagnetic activity, simple diurnal variations in temperature and heating would significantly extend each refilling stage and decrease the refilling rates to values more comparable to those listed in Table 3 of Gallagher et al. (2021). This is assuming that the current temperatures and heating rates are reflective of dayside conditions. Even a simple step function for dayside and nightside boundary temperatures and heating rates should improve the refilling stage lengths and rates and begin to indicate how impactful these variabilities are. Afterward, it would be ideal to formulate a more sophisticated variability of boundary temperatures and heating rates reflective of how the various heating mechanisms at play vary with space and time.

Another goal for improving the physicality of the model is increasing its compatibility with observations, including those of specific events. Coupling to a preexisting plasmasphere-ionosphere model such as the Ionosphere-Plasmasphere-Electrodynamics (IPE) model (Maruyama et al., 2016) would supply more accurate and time-dependent boundary conditions. Therefore, the resulting refilling produced by the model should exhibit features more similar to those seen in observation. Extracting boundary conditions and heating rates for a particular storm would permit a more in-depth comparison with the respective refilling event to suggest room for improvement in both the model and observational methods. Eventually, the model will be extrapolated to 3D to achieve a global perspective of plasmasphere refilling.

## 5. Conclusions

This study extends the multi-ion, two-stream, hydrodynamic Flux-Corrected Transport (FCT) plasmasphere refilling model by incorporating the electron energy equation, enabling spatially and temporally varying plasma temperatures during plasmasphere refilling. This addition introduced a new physical behavior of two-stage refilling that could not be captured under the assumption of a constant temperature along the flux tube. These results demonstrate that self-consistent temperatures are a key component of refilling dynamics and are necessary to accurately reproduce the two-stage process. The temperature-coupled model offers a more complete physical description of plasmasphere refilling than earlier constant-temperature formulations, and it provides a framework for investigating variability across geomagnetic storm events.

Using this model to analyze the plasma contents of the flux tube throughout refilling revealed the impact of individual ions on the two-stage process. The greater presence of $O^+$ in early-time refilling supports the notion suggested by previous literature that heavy ions play a critical role at the start of refilling. As early-time transitioned to late-time refilling, the ambipolar electric field appeared to couple the $H^+$ and $He^+$ concentrations, which could only possibly be observed with the spatiotemporal temperature variability introduced by this study. The relative importance of each ion species during different segments of refilling were further revealed by simulating refilling events with various initial ion concentration profiles, including those reflecting seasonal variations.

Future work should incorporate spatially and temporally varying heating rates, cooling processes, and coupling to global ionosphere–thermosphere and ring-current models. Expanding the model to three dimensions would also allow investigation of cross-field plasma transport and magnetic local time effects, providing further insight into the complex variability of plasmasphere recovery following geomagnetic storms.

## Conflict of Interest

The authors declare no conflicts of interest relevant to this study.

## Availability Statement

The model data used to produce each figure is provided on CU Scholar at https://doi.org/10.25810/9vf2-1m93 (Fitzpatrick et al., 2026).

**Acknowledgments**
We would like to thank R.W. Schunk for his support and contributions in developing the code, and X. Chu, T. Bishop, and L. Blum for their valuable discussions in preparing the manuscript. We would like to acknowledge the University of Colorado Boulder, the Laboratory for Atmospheric and Space Physics, and the Charles A Barth Scholarship in Space Research for supporting this work. We would also like to thank Grants NASA 80NSSC20K1817, 80NSSC22K1023, 80NSSC23M0192, 80GSFC23CA004, 80NSSC20K1351, 80NSSC25K7762, NSF Grant AGS-2412296, and AFOSR FA9550-24-1-0013. This work utilized the Alpine high-performance computing resource at the University of Colorado Boulder. Alpine is jointly funded by the University of Colorado Boulder, the University of Colorado Anschutz, Colorado State University, and the National Science Foundation (award 2201538).